\documentclass[twocolumn,pre]{revtex4}

\usepackage{dcolumn}
\usepackage{amsmath}

\usepackage{graphicx}
\usepackage{subfigure}

\newcommand{\e}{\mathrm{e}}

\newcommand{\defn}{\textit}
\newcommand{\Ord}{\mathrm{O}}
\newcommand{\mat}{\mathbf}
\renewcommand{\vec}{\mathbf}

\newcommand\pin{p_\textrm{in}}
\newcommand\pout{p_\textrm{out}}
\newcommand\win{\omega_\textrm{in}}
\newcommand\wout{\omega_\textrm{out}}
\newcommand\cin{c_\textrm{in}}
\newcommand\cout{c_\textrm{out}}
\newcommand\emin{m_\textrm{in}}
\newcommand\emout{m_\textrm{out}}

\begin{document}

\title{Community detection and graph partitioning}
\author{M. E. J. Newman}
\affiliation{Department of Physics and Center for the Study of Complex
  Systems, University of Michigan, Ann Arbor, MI 48109}

\begin{abstract}
  Many methods have been proposed for community detection in networks.
  Some of the most promising are methods based on statistical inference,
  which rest on solid mathematical foundations and return excellent results
  in practice.  In this paper we show that two of the most widely used
  inference methods can be mapped directly onto versions of the standard
  minimum-cut graph partitioning problem, which allows us to apply any of
  the many well-understood partitioning algorithms to the solution of
  community detection problems.  We illustrate the approach by adapting the
  Laplacian spectral partitioning method to perform community inference,
  testing the resulting algorithm on a range of examples, including
  computer-generated and real-world networks.  Both the quality of the
  results and the running time rival the best previous methods.
\end{abstract}

\maketitle

\section{Introduction}
The problem of community detection in networks has received wide
attention~\cite{GN02,Fortunato10}.  It has proved to be a problem of
remarkable subtlety, computationally challenging and with deep connections
to other areas of research including machine learning, signal processing,
and spin-glass theory.  A large number of algorithmic approaches to the
problem have been considered, but interest in recent years has focused
particularly on statistical inference methods~\cite{ABFX08,CMN08,BC09},
partly because they give excellent results, but also because they are
mathematically principled and, at least in some cases, provably
optimal~\cite{BC09,DKMZ11a}.

In this paper we study two of the most fundamental community inference
methods, based on the so-called stochastic block model or its
degree-corrected variant~\cite{KN11a}.  We show that it is possible to map
both methods onto the well-known minimum-cut graph partitioning problem,
which allows us to adapt any of the large number of available methods for
graph partitioning to solve the community detection problem.  As an
example, we apply the Laplacian spectral partitioning method of
Fiedler~\cite{Fiedler73,PSL90} to derive a community detection method
competitive with the best currently available algorithms in terms of both
speed and quality of results.

\section{Likelihood maximization for the stochastic block model}
The first method we consider is based on the \defn{stochastic block model},
sometimes also called the \defn{planted partition model}, a well studied
model of community structure in networks~\cite{KN11a,CK01}.  This model
supposes a network of~$n$ vertices divided into some number of groups or
communities, with different probabilities for connections within and
between groups.  We will here focus on the simplest case of just two groups
(of any size, not necessarily equal).  In the commonest version of the
model edges are placed independently at random between vertex pairs with
probability~$\pin$ for pairs in the same group and $\pout$ for pairs in
different groups.  In this paper we use the slightly different Poisson
version of the model described in~\cite{KN11a}, in which we place between
each pair of vertices a Poisson-distributed number of edges with mean
$\win$ for pairs in the same group and $\wout$ for pairs in different
groups.  In essentially all real-world networks the fraction of possible
edges that are actually present in the network is extremely small (usually
modeled as vanishing in the large-$n$ limit), in which case the two
versions of the model become indistinguishable, but the Poisson version is
preferred because its analysis is more straightforward.

At its heart, the statistical inference of community structure is a matter
of answering the following question: if we assume an observed network is
generated according to our model, what then must the parameters of that
model have been?  In other words, what were the values of $\win$
and~$\wout$ used to generate the network and, more importantly, which
vertices fell in which groups?  Even though the model is probably not a
good representation of the process by which most real-world networks are
generated, the answer to this question often gives a surprisingly good
estimate of the true community structure.

To answer the question, we make use of a maximum likelihood method.  Let us
label the two groups or communities in our model group~1 and group~2, and
denote by~$g_i$ the group to which vertex~$i$ belongs.  The edges in the
network will be represented by an adjacency matrix having elements
\begin{equation}
A_{ij} = \biggl\lbrace\begin{array}{ll}
  1 & \quad\mbox{if there is an edge between vertices $i,j$,} \\
  0 & \quad\mbox{otherwise.}
\end{array}
\end{equation}
Then the likelihood of generating a particular network or graph~$G$, given
the complete set of group memberships, which we'll denote by the
shorthand~$g$, and the Poisson parameters, which we'll denote by~$\omega$,
is
\begin{equation}
P(G|g,\omega) = \prod_{i<j} {\omega_{ij}^{A_{ij}}\over A_{ij}!}
  \e^{-\omega_{ij}},
\label{eq:likelihood}
\end{equation}
where $\omega_{ij}$ denotes the expected number of edges between
vertices~$i$ and~$j$---either $\win$ or~$\wout$, depending on whether the
vertices are in the same or different groups.  We are assuming there are no
self-edges in the network---edges that connect vertices to themselves---so
$A_{ii}=0$ for all~$i$.

Given the likelihood, one can maximize it to find the most likely values of
the group labels and parameters, which can be done in a number of different
ways.  In Ref.~\cite{KN11a}, for example, the likelihood was maximized
first with respect to the parameters~$\win$ and~$\wout$ by differentiation.
Applying this method to Eq.~\eqref{eq:likelihood} gives most likely values
of
\begin{equation}
\win = {2\emin\over n_1^2+n_2^2}, \qquad
\wout = {\emout\over n_1n_2},
\end{equation}
where $\emin$ and $\emout$ are the observed numbers of edges within and
between groups respectively for a given candidate division of the network,
and $n_1$ and $n_2$ are the numbers of vertices in each group.
Substituting these values back into Eq.~\eqref{eq:likelihood} gives the
\defn{profile likelihood}, which depends on the group labels only.  In
fact, one typically quotes not the profile likelihood itself but its
logarithm, which is easier to work with.  Neglecting an unimportant
additive constant, the log of the profile likelihood for the present model
is
\begin{equation}
\mathcal{Q} = \emin \ln {2\emin\over n_1^2+n_2^2}
              + \emout \ln {\emout\over n_1 n_2}.
\label{eq:profile}
\end{equation}
The communities can now be identified by maximizing this quantity over all
possible assignments of the vertices to the groups.  This is still a hard
task, however.  There are an exponentially large number of possible
assignments, so an exhaustive search through all of them is unfeasible for
all but the smallest of networks.  One can apply standard heuristics like
simulated annealing to the problem, but in this paper we take a different
approach.

In the calculation above, the likelihood is maximized over $\omega$ first,
for fixed group assignments, then over the group assignments.  But we can
also take the reverse approach, maximizing first over the group
assignments, for given~$\omega$, and then over~$\omega$ at the end.  This
approach is attractive for two reasons.  First, as we will show, the
problem of maximizing with respect to the group assignment when $\omega$ is
given is equivalent to the standard problem of minimum-cut graph
partitioning, a problem for which many excellent heuristics are already
available.  Second, after maximizing with respect to the group assignments
the remaining problem of maximizing with respect to~$\omega$ is a
one-parameter optimization that can be solved trivially.  The net result is
that the problem of maximum-likelihood community detection is reduced to
one of performing a well-understood task---graph partitioning---plus one
undemanding extra step.  The resulting algorithm is fast and, as we will
see, gives good results.

So consider the problem of maximizing the likelihood,
Eq.~\eqref{eq:likelihood}, with respect to the group labels~$g_i$, for
given values of the parameters~$\win$ and~$\wout$.  We will actually
maximize the logarithm~$\mathcal{L}$ of the likelihood,
\begin{equation}
  \mathcal{L} = \ln P(G|g,\omega)
  = \sum_{i<j} \bigl[ A_{ij} \ln \omega_{ij} - \omega_{ij}
  - \ln A_{ij}! \bigr],
\label{eq:logl1}
\end{equation}
which gives the same result but is usually easier.

To proceed we write~$\omega_{ij}$ and $\ln\omega_{ij}$ as
\begin{align}
\omega_{ij} &= \delta_{g_ig_j} \win + (1-\delta_{g_ig_j})\,\wout, \\
\ln\omega_{ij} &= \delta_{g_ig_j} \ln\win + (1-\delta_{g_ig_j}) \ln\wout,
\end{align}
where $\delta_{ij}$ is the Kronecker delta.  Substituting these into
Eq.~\eqref{eq:logl1} and dropping overall additive and multiplicative
constants, which have no effect on the position of the maximum, the
log-likelihood can be rearranged to read
\begin{equation}
\mathcal{L} = \sum_{i<j} ( 1 - \delta_{g_ig_j} )
              ( \gamma - A_{ij} ),
\label{eq:logl}
\end{equation}
where
\begin{equation}
\gamma = {\win-\wout\over\ln\win-\ln\wout},
\label{eq:alpha}
\end{equation}
which is positive whenever $\win>\wout$, meaning we have traditional
community structure in our network.  (It is possible to repeat the
calculations for the case $\win<\wout$ and derive methods for detecting
such structure as well, although we will not do that here.)

The quantity $\sum_{i<j} ( 1 - \delta_{g_ig_j} ) A_{ij}$ is the \defn{cut
  size} of the network partition represented by our two communities,
i.e.,~the number of edges connecting vertices in different communities,
which we previously denoted~$\emout$, and
\begin{equation}
\sum_{i<j} ( 1 - \delta_{g_ig_j} ) = n_1 n_2,
\end{equation}
where as previously $n_1$ and $n_2$ are the numbers of vertices in
communities~1 and~2.  Thus we can also write the log-likelihood in the form
\begin{equation}
\mathcal{L} = -\emout + \gamma n_1 n_2.
\label{eq:diffcut}
\end{equation}
The maximization of this log-likelihood corresponds to the minimization of
the cut size, with an additional penalty term~$\gamma n_1 n_2$ that favors
groups of equal size.  This is similar, though not identical, to the
so-called \defn{ratio cut} problem, in which one minimizes the ratio
$\emout/n_1n_2$, which also favors groups of equal size, although the
nature of the penalty for unbalanced groups is different.

The catch with maximizing Eq.~\eqref{eq:diffcut} is that we don't know the
value of~$\gamma$, which depends on the unknown quantities $\win$
and~$\wout$ via Eq.~\eqref{eq:alpha}, but we can get around this problem by
the following trick.  We first perform a limited maximization
of~\eqref{eq:diffcut} in which the sizes $n_1$ and~$n_2$ of the groups are
held fixed at some values that we choose.  This means that the term $\gamma
n_1n_2$ is a constant and hence drops out of the problem and we are left
maximizing only~$-\emout$, or equivalently minimizing the
cut-size~$\emout$.  This problem is now precisely the standard minimum-cut
problem of graph partitioning---the minimization of the cut size for
divisions of a graph into groups of given sizes.

There are $n+1$ possible choices of the sizes of the two groups, ranging
from putting all vertices in group~1 to all vertices in group~2, and
everything in between.  If we solve the minimum-cut problem for each of
these $n+1$ choices we get a set of $n+1$ solutions and we know that one of
these must be the solution to our overall maximum likelihood problem.  It
remains only to work out which one.  But choosing between them is easy,
since we know that the true maximum also maximizes the profile likelihood,
Eq.~\eqref{eq:profile}.  So we can simply calculate the profile likelihood
for each solution in turn and find the one that gives the largest result.

In effect, our approach narrows the exponentially large pool of candidate
divisions of the network to a one-parameter family of just $n+1$ solutions
(parametrized by group size), from which it is straightforward to pick the
overall winner by exhaustive search.  Moreover, the individual candidate
solutions are all themselves solutions of the standard minimum-cut
partitioning problem, a problem that has been well studied for many years
and about which a great deal is known~\cite{Elsner97,Fjallstrom98}.
Although partitioning problems are, in general, hard to solve exactly,
there exist many heuristics that give good answers in practical situations.
The approach developed here allows us to apply any of these heuristics
directly to the maximum-likelihood community detection problem.

\subsection{Spectral algorithm}
As an example of this approach, we demonstrate a fast and simple spectral
algorithm based on the Laplacian spectral bisection method for graph
partitioning introduced by Fiedler~\cite{Fiedler73,PSL90}.  A description
of this method can be found, for example, in~\cite{Newman10}, where it is
shown that a good approximation to the minimum-cut division of a network
into two parts of specified sizes can be found by calculating the
\defn{Fiedler vector}, which is the eigenvector of the graph Laplacian
matrix~$\mat{L}$ corresponding to the second smallest eigenvalue.  (The
graph Laplacian is the $n\times n$ symmetric matrix $\mat{L} = \mat{D} -
\mat{A}$, where $\mat{A}$ is the adjacency matrix and $\mat{D}$ is the
$n\times n$ diagonal matrix with~$D_{ii}$ equal to the degree of
vertex~$i$.)  Having calculated the Fiedler vector one divides the network
into groups of the required sizes~$n_1$ and $n_2$ by inspecting the vector
elements and assigning the~$n_1$ vertices with the largest (most positive)
elements to group~1 and the rest to group~2.  Although the method gives
only an approximation to the global minimum-cut division, practical
experience (and some rigorous results) show that it gives good answers
under commonly occurring conditions~\cite{PSL90}.

A nice feature of this approach is that, in a single calculation, it gives
us the entire one-parameter family of minimum-cut divisions of the network.
We need calculate the Fiedler vector only once, sort its elements in
decreasing order, then cut them into two groups in each of the $n+1$
possible ways and calculate the profile likelihood for the resulting
divisions of the network.  The one with the highest score is (an
approximation to) the maximum-likelihood community division of the network.

\subsection{Degree-corrected block model}
These developments are for the standard stochastic block model.  As shown
in Ref.~\cite{KN11a}, however, the standard block model gives poor results
when applied to most real-world networks because the model fails to take
into account the broad degree distribution such networks possess.  This
problem can be fixed by a relatively simple modification of the model in
which the expected number~$\omega_{ij}$ of edges between vertices~$i$
and~$j$ is replaced by $k_ik_j\omega_{ij}$ where $k_i$ is the degree of
vertex~$i$ and $\omega_{ij}$ again depends only on which groups the
vertices~$i$ and~$j$ belong to.  All the developments for the standard
block model above generalize in straightforward fashion to this
``degree-corrected'' model.  The log-likelihood and log-profile likelihood
become
\begin{equation}
\mathcal{L} = -\emout + \gamma \kappa_1 \kappa_2, \quad
\mathcal{Q} = \emin \ln {2\emin\over\kappa_1^2+\kappa_2^2}
              + \emout \ln {\emout\over\kappa_1\kappa_2},
\label{eq:likelydc}
\end{equation}
where $\kappa_1$ and $\kappa_2$ are the sums of the degrees of the vertices
in the two groups.  In other words, the expressions are identical to those
for the uncorrected model except for the replacement of the group
sizes~$n_1,n_2$ by~$\kappa_1,\kappa_2$.

The maximization of~$\mathcal{L}$ is thus once again reduced to a
generalized minimum-cut partitioning problem, with a penalty term
proportional to $\kappa_1\kappa_2$, which again favors balanced groups.
Although we don't know the value of~$\gamma$, we can reduce the problem to
a variant of the minimum-cut problem by the equivalent of our previous
approach, holding~$\kappa_1$ and~$\kappa_2$ constant.  And again we can
derive a spectral algorithm for this problem based on the graph Laplacian.
By a derivation analogous to that for the standard spectral method we can
show that a good approximation to the problem of minimum-cut partitioning
with fixed~$\kappa_1,\kappa_2$ (as opposed to fixed~$n_1,n_2$) is given not
by the second eigenvector of~$\mat{L}$ but by the second eigenvector of the
generalized eigensystem $\mat{L}\vec{v} = \lambda\mat{D}\vec{v}$, where, as
previously, $\mat{D}$~is the diagonal matrix of vertex degrees.  Once again
we calculate the vector and split the vertices into two groups according to
the sizes of their corresponding vector elements and once again this gives
us a one-parameter family of $n+1$ candidate solutions from which we can
choose an overall winner by finding the one with the highest profile
likelihood, Eq.~\eqref{eq:likelydc}.

\begin{figure}
\begin{center}
\includegraphics[width=\columnwidth]{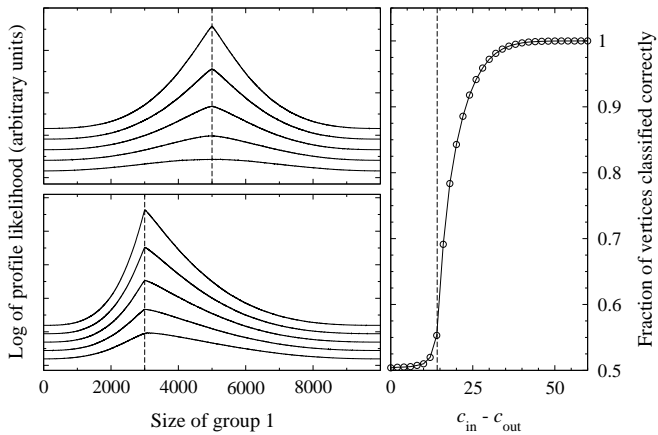}
\end{center}
\caption{(a)~Profile likelihood as a function of group size for candidate
  solutions generated from the spectral method for single network of
  $n=10\,000$ vertices, generated using the standard (uncorrected)
  stochastic block model with equal group sizes of 5000 vertices each and a
  range of strengths of the community structure.  Defining $\cin=n\win$,
  $\cout=n\wout$, the curves are (top to bottom) $\cin=80$, 75, 70, 65,
  and~60, and $\cout=100-\cin$.  The dashed vertical line indicates the
  true size of the planted communities.  The curves have been displaced
  from one another vertically for clarity.  The vertical axis units are
  arbitrary because additive and multiplicative constants have been
  neglected in the definition of the log-likelihood.  (b)~Profile
  likelihoods for the same parameter values but unequal groups of size
  $3000$ and~$7000$.  (c)~The average fraction of vertices classified
  correctly for networks of $10\,000$ vertices each and two equally sized
  groups.  Each point is an average over 100 networks.  Statistical errors
  are smaller than the points in all cases.  The vertical dashed line
  indicates the position of the ``detectability threshold'' at which
  community structure becomes formally
  undetectable~\cite{RL08,DKMZ11a,HRN12,NN12}.}
\label{fig:synthetic}
\end{figure}

\section{Results}
We have tested this method on a variety of networks, and in practice it
appears to work well.  Figure~\ref{fig:synthetic} shows results from tests
on a large group of synthetic (i.e.,~computer-generated) networks.  These
networks were themselves generated using the standard stochastic block
model (which is commonly used as a benchmark for community
detection~\cite{CK01,GN02}).  The two left panels in the figure show the
value of the profile likelihood for the families of $n+1$ candidate
solutions generated by the spectral calculation for networks with two
equally sized groups (top) and with unequal groups (bottom).  In each case
there is a clear peak in the profile likelihood at the correct group sizes,
suggesting that the algorithm has correctly identified the group membership
of most vertices.  The third panel in Fig.~\ref{fig:synthetic} tests this
conclusion by calculating the fraction of correctly identified vertices as
a function of the strength of the community structure for equally sized
groups (which is the most difficult case).  As the figure shows, the
algorithm correctly identifies most vertices over a large portion of the
parameter space.  The vertical dashed line represents the ``detectability
threshold'' identified by previous
authors~\cite{RL08,DKMZ11a,HRN12,NN12,MNS12}, below which it is believed
that every method of community detection must fail.  Our algorithm fails
below this point also, but appears to work well essentially all the way
down to the transition, and there are reasons to believe this result to be
exact, at least for networks that are not too sparse~\cite{NN12}.

\begin{figure}
\begin{center}
\includegraphics[width=6.5cm]{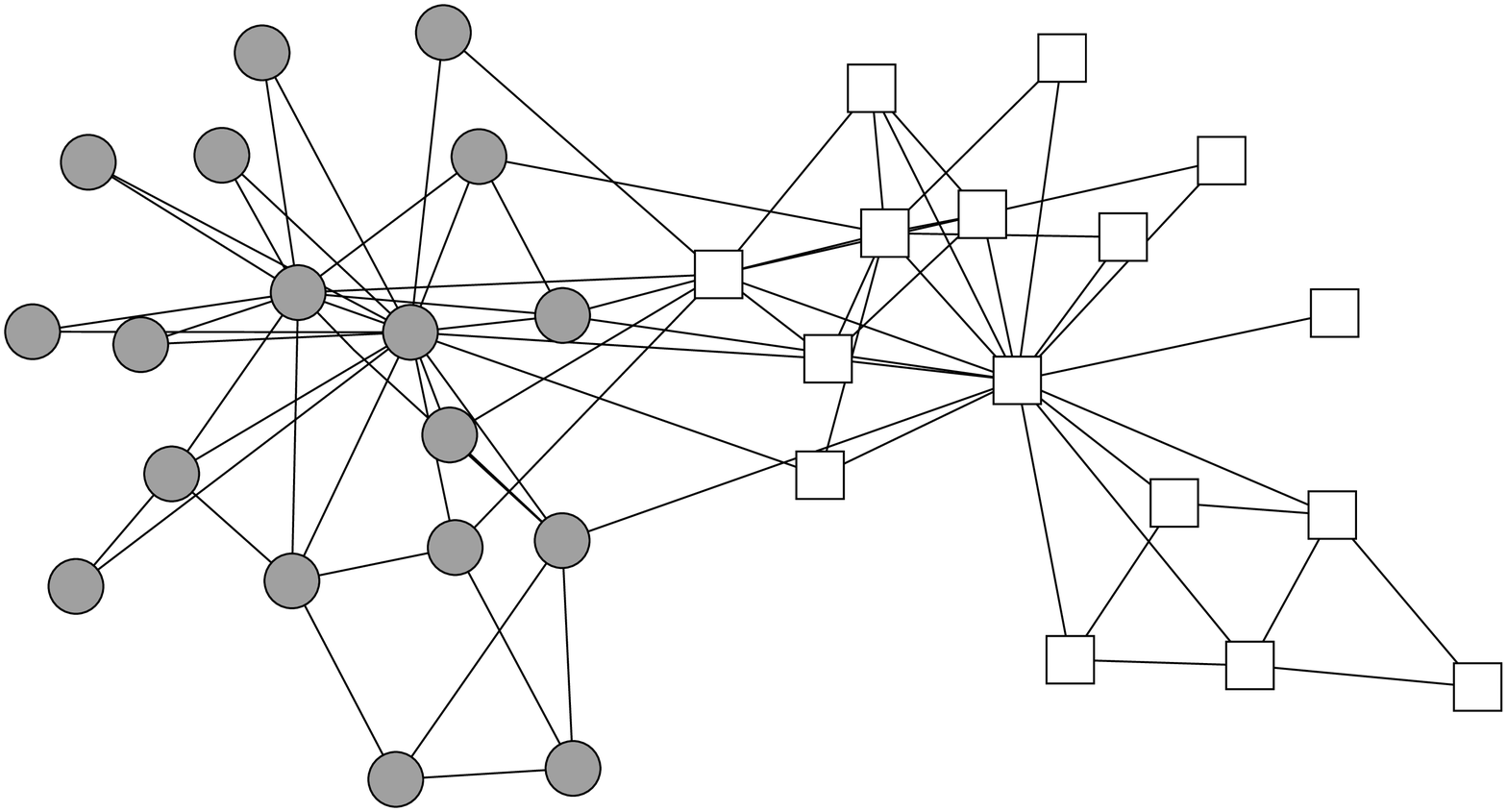} \\
{\ }\\
\includegraphics[width=7.5cm]{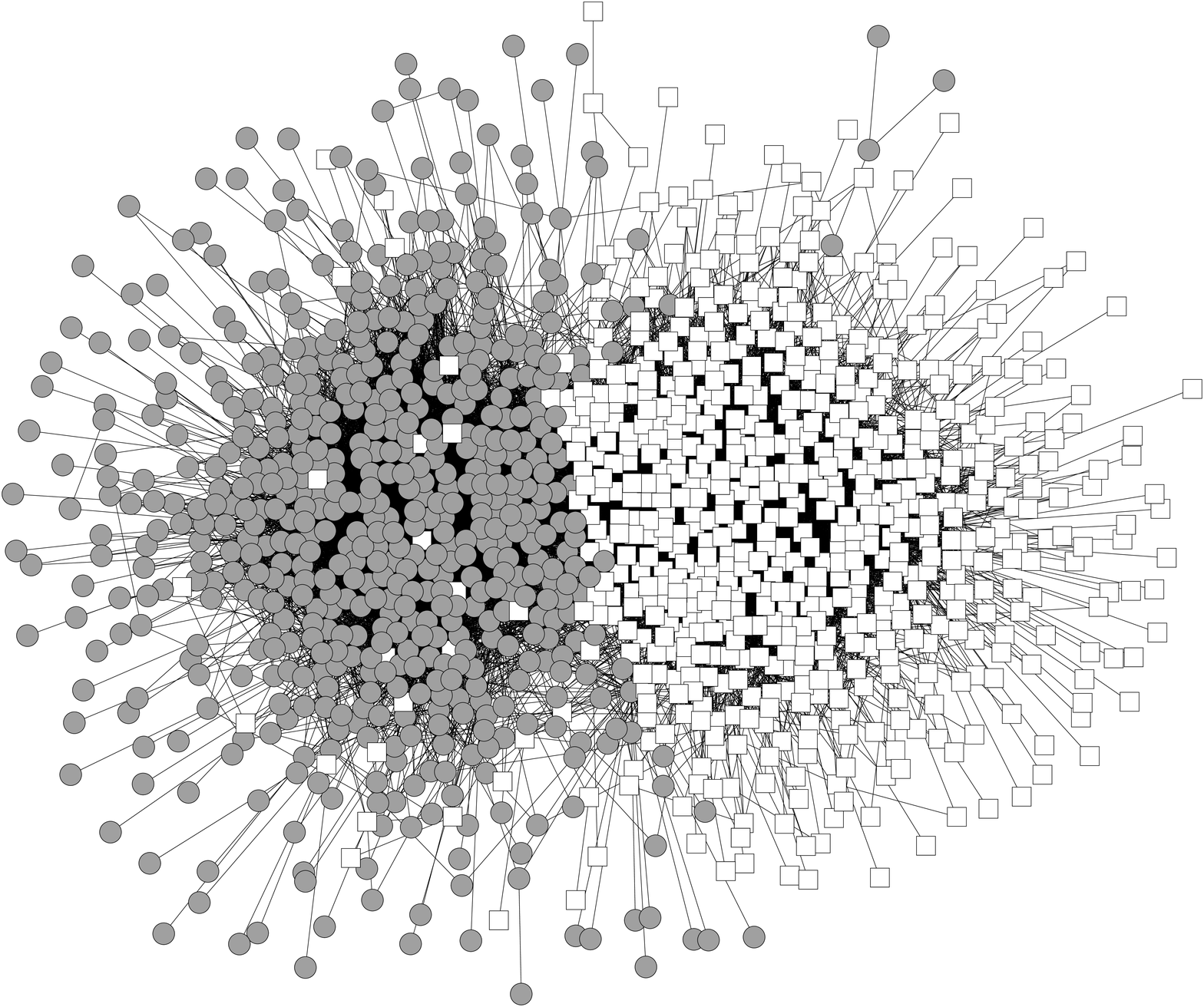}
\end{center}
\caption{The division into two groups of two well-known networks from the
  literature.  Top: the karate club network of Zachary~\cite{Zachary77}.
  Bottom: the network of political blogs compiled by Adamic and
  Glance~\cite{AG05}.  Vertices colors and shapes indicate the group
  membership and both divisions are qualitatively similar to the accepted
  ones.}
\label{fig:results}
\end{figure}

Figure~\ref{fig:results} shows the results of applications of the algorithm
to two well-studied real-world networks, Zachary's ``karate club''
network~\cite{Zachary77} and Adamic and Glance's network of political
blogs~\cite{AG05}.  Both are known to have pronounced community structure
and the divisions found by our spectral algorithm mirror closely the
accepted communities in both cases.

In addition to being effective, the algorithm is also fast.  The
computation of the eigenvector can be done using, for instance, the Lanczos
method, an iterative method which takes time~$\Ord(m)$ per iteration, where
$m$ is the number of edges in the network.  The number of iterations
required is typically small, although the exact number is not known in
general.  The search for the division that maximizes the profile likelihood
can also be achieved in $\Ord(m)$ time.  Of the $n+1$ different divisions
of the network that must be considered, each one differs from the previous
one by the movement of just a single vertex from one group to the other.
The movement of vertex~$i$ between groups causes the quantities appearing
in Eq.~\eqref{eq:likelydc} to change according to
\begin{align}
& \kappa_1 \to \kappa_1 - k_i,\quad \kappa_2 \to \kappa_2 + k_i, \\
& \emin \to \emin - \Delta m,\quad \emout \to \emout + \Delta m,
\end{align}
where $\Delta m$ equals the number of edges between~$i$ and vertices in
group~1 minus the number between~$i$ and vertices in group~2.  These
quantities and the resulting change in the profile likelihood can be
calculated in time proportional to the degree of the vertex and hence all
$n$ vertices can be moved in time proportional to the sum of all degrees in
the network, which is equal to~$2m$.  Thus, to leading order, the total
running time of the algorithm goes as~$m$ times the number of Lanczos
iterations, the latter typically being small, and in practice the method is
about as fast as the best competing algorithms.

\section{Conclusions}
In this paper we have shown that the widely-studied maximum likelihood
method for community detection in networks can be reduced to a search
through a small family of candidate solutions, each of which is itself the
solution to a minimum-cut graph partitioning problem, which is a well
studied problem about which much is known.  This mapping allows us to use
trusted partitioning heuristics to solve the community detection problem.
As an example we have adapted the method of Laplacian spectral partitioning
to derive a spectral likelihood maximization algorithm and tested its
performance on both synthetic and real-world networks.  In terms of both
accuracy and speed we find the algorithm to be competitive with the best
current methods.

A number of extensions of our approach would be possible, including
extensions with more general forms for the parameters~$\omega$, such as
different values of $\win$ and $\wout$ for different groups, or extensions
to more than two groups, but we leave these for future work.

\begin{acknowledgments}
  The author would like to thank Charlie Doering, Tammy Kolda, and Raj Rao
  Nadakuditi for useful conversations and Lada Adamic for providing the
  data for the network of political blogs.  This work was funded in part by
  the National Science Foundation under grant DMS--1107796 and by the Air
  Force Office of Scientific Research (AFOSR) and the Defense Advanced
  Research Projects Agency (DARPA) under grant FA9550--12--1--0432.
\end{acknowledgments}

\end{document}